%% file: main.tex
\documentclass[journal]{IEEEtran}

\ifCLASSINFOpdf
\else
\fi

\PassOptionsToPackage{dvipsnames}{xcolor}

\usepackage{multirow, booktabs}
\usepackage{amsmath}
\usepackage{amsfonts}
\usepackage{bm}
\usepackage{enumitem}
 \usepackage{graphicx}
 \usepackage{pifont}
 \usepackage{multicol}
 \usepackage{caption}
\usepackage{subcaption}
\usepackage[numbers,sort]{natbib}

\usepackage{url}
\usepackage{xfrac}
\usepackage{wrapfig}
\usepackage{multirow, booktabs}
\usepackage{tabularx}
\usepackage{xfrac}
\usepackage{xcolor}
\usepackage[colorlinks=true,citecolor=NavyBlue,linkcolor=ForestGreen]{hyperref}%
\usepackage{url}
\usepackage{epigraph} 
\usepackage{titlesec}
\usepackage{mathtools}
\newcommand{\defeq}{\vcentcolon=}
\titleformat{\paragraph}[runin]{\normalfont\normalsize\bfseries}{}{}{}
\titlespacing{\paragraph}{0pt}{3.5ex plus 1ex minus .2ex}{2.3ex plus .2ex}
\newcolumntype{R}{>{\raggedleft\arraybackslash}X}
\newcolumntype{L}{>{\raggedright\arraybackslash}X}
\newcolumntype{C}{>{\centering\arraybackslash}X}
\newcommand{\cmark}{\ding{51}}
\newcommand{\xmark}{\ding{55}}

\newcommand{\cd}[1]{}
\newcommand{\sw}[1]{}
\newcommand{\nb}[1]{}
\newcommand{\todo}[1]{}
\newcommand{\slwu}[1]{}

\newcommand{\pcite}[1]{\citep{#1}}
\newcommand{\globl}{global}
\newcommand{\timevar}{time-varying}
\newcommand{\eref}[1]{(\ref{#1})}

\newcommand{\sref}[1]{Section~\ref{#1}}
\newcommand{\fref}[1]{Fig.~\ref{#1}}

\newcommand{\extracted}{extracted}
\newcommand{\expected}{created}
\newcommand{\Extracted}{Extracted}
\newcommand{\Expected}{Created}
\newcommand{\strict}{strict}

\newcommand{\wave}{\bm{w}}

\newcommand{\spec}{\bm{s}}

\begin{document}
%
\title{\textsc{Music ControlNet}: Multiple Time-varying Controls for Music Generation}
%
%
%
\author{Shih-Lun~Wu$^{1, 2*}$,
        Chris~Donahue$^{1}$,
        Shinji~Watanabe$^{1}$,
        and Nicholas~J.~Bryan$^{2}$ \\
        \vspace{2mm}
        $^{1}$ School of Computer Science, Carnegie Mellon University \;\;\; $^{2}$ Adobe Research \\
        \thanks{$^*$Work done during Shih-Lun's internship at Adobe Research. Correspondence should be addressed to Shih-Lun Wu and Nicholas J.~Bryan at \texttt{shihlunw@cs.cmu.edu} and \texttt{njb@ieee.org}, respectively.}
}
\maketitle
\input{sections/abstract}

\input{sections/introduction}

\input{sections/background}

\input{sections/method}

\input{sections/experimental_design}

\input{sections/results}

\input{sections/related}

\input{sections/conclusion}

\input{sections/ethics}

\section{Acknowledgements}
Thank you to Ge Zhu, Juan-Pablo Caceres, Zhiyao Duan, and Nicholas J. Bryan for sharing their high-fidelity vocoder used for the demo video (citation coming soon).

\bibliography{refs}
\bibliographystyle{IEEEtran}

\end{document}

%% file: sections/abstract.tex
\begin{abstract}
Text-to-music generation models are now capable of generating high-quality music audio in broad styles. 
However, text control
is primarily 
suitable 
for the manipulation of \emph{\globl} musical attributes like genre, mood, and 
tempo, 
and is less suitable for precise control over \emph{\timevar{}} attributes such as the positions of beats in time or the changing dynamics of the music. 
We propose Music ControlNet, a diffusion-based music generation model that offers multiple precise, \timevar{} controls over generated audio. To imbue text-to-music models with \timevar{} control, we propose an approach analogous to pixel-wise control of the image-domain ControlNet method. 
Specifically, we extract controls from training audio yielding 
paired data, and fine-tune a diffusion-based conditional generative model over audio spectrograms given melody, dynamics, and rhythm controls. 
While the image-domain Uni-ControlNet method already allows generation with any subset of controls, we devise a new masking strategy at training to allow creators to input controls that are only partially specified in time. 
We evaluate both on controls \extracted{} from audio 
and controls we expect creators to provide, 
demonstrating that we can generate
realistic 
music that corresponds to control inputs in both settings. 
While few comparable music generation models exist, 
we benchmark against 
MusicGen, a recent model that accepts text and melody input, 
and show that our model generates music that is 
$49$\% more faithful to 
input melodies despite 
having $35$x fewer parameters, training on $11$x less data, and enabling two additional forms of \timevar{} control. 
Sound examples can be found at \url{https://MusicControlNet.github.io/web/}.
\end{abstract}

\begin{IEEEkeywords}
music generation, controllable generative modeling, diffusion
\end{IEEEkeywords}

%% file: sections/introduction.tex
\section{Introduction}\label{sec:intro}

One of the pillars of musical expression is the communication of high-level ideas and emotions through precise manipulation of lower-level attributes like notes, dynamics, and rhythms. Recently, there has been an explosion of interest in training text-to-music generative models that allow creators to directly convert high-level intent (expressed as text) into music audio~\citep{dhariwal2020jukebox, agostinelli2023musiclm, huang2023noise2music, liu2023audioldm, copet2023simple}. These models suggest an exciting new paradigm of musical expression wherein  creators can instantaneously generate realistic music without the need to 
  write a melody, 
  specify meter and rhythm, 
  or 
  orchestrate instruments. 
 However, while dramatically more efficient, this new paradigm ignores more conventional forms of musical expression rooted in the manipulation of lower-level attributes, 
 limiting the ability to express precise musical intent or leverage models in existing creative workflows.

\begin{figure*}[t!]
    \centering
   \begin{center}
    \includegraphics[width=0.92\textwidth,trim={0mm, 2mm, 0, 2mm},clip]{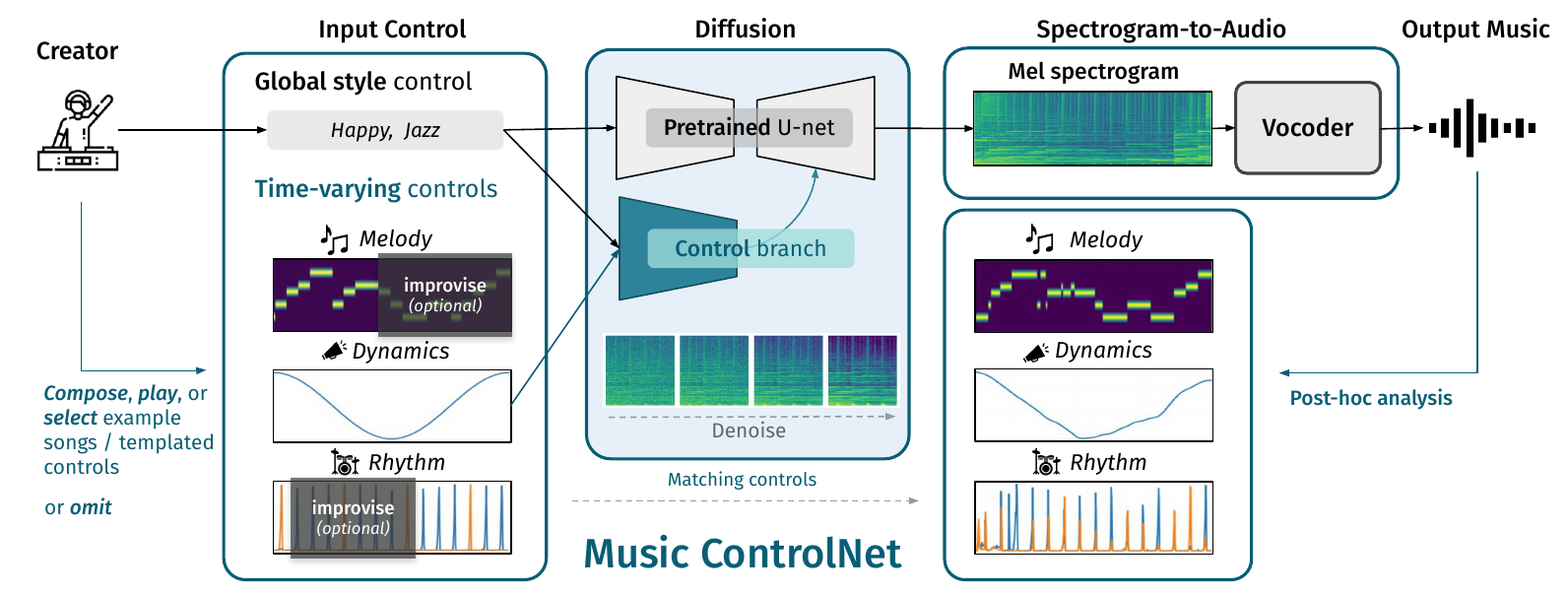}
    \end{center}
    \caption{Music ControlNet overview.
    Our model accepts as input \globl{} genre and mood text control, 
    alongside 
    any combinations of precise, \timevar{} melody, dynamics, and rhythm controls. 
    The controls can each be fully or partially specified in time,
    the latter of which signals the model to musically improvise.
    Music that adheres to input controls is generated using a diffusion model that outputs an image-like representation of music (a Mel spectrogram), which is then rendered as audio using a
    vocoder. 
    Music ControlNet empowers creators to blend text and musical controls in their creative process with a straightforward pipeline.
    }
    \label{fig:headline}
\end{figure*}

There are two primary obstacles for adding precise control to text-based music generation methods. 
Firstly, relative to symbolic music representations like scores, 
text is a cumbersome interface for conveying precise musical attributes that vary over time. 
Verbose and mundane text descriptions may be needed to precisely represent even the first note of a musical score 
e.g.,~``the song starts at $80$ beats per minute with a quarter note on middle C played mezzo-forte on the saxophone''. 
The second obstacle is an empirical one---text-to-music models tend to faithfully interpret \emph{\globl} stylistic attributes (e.g.,~genre and mood) from text, but struggle to interpret text descriptions of precise musical attributes (e.g.,~notes or rhythms). 
This is perhaps a consequence of the relative scarcity of precise descriptions in the training data.

A potential solution to the musical imprecision of natural language is the incorporation of \emph{\timevar{}} controls into music generation. 
For example, one body of work looks at synthesizing music audio from \timevar{} symbolic music representations like MIDI~\citep{hawthorne2018enabling,hawthorne2022multi}, however this approach offers a particularly \strict{} form of control requiring users to compose entire pieces of music beforehand. Such approaches are more similar to typical music composition processes and do not take full advantage of recent text-to-music methods.
Another body of work on musical style transfer~\citep{mor2018universal,huang2018timbretron,engel2020ddsp,caillon2021rave,wu2021midi, steinmetz2022style} seeks to transform recordings from one \emph{style} (e.g., genre, musical ensemble, or mood) to another while preserving the underlying composition content. 
However, 
a majority of these approaches require training an individual model per style, 
as opposed to the flexibility of using text to control style in a single model.

In this work, we propose Music ControlNet, a diffusion-based music generation model that offers multiple \timevar{} controls over the melody, dynamics, and rhythm of generated audio, in addition to global text-based style 
control as shown in~\fref{fig:headline}.  
To incorporate 
such \timevar{} controls,
we adapt recent work on image generation with spatial control, namely, ControlNet~\pcite{zhang2023adding} and Uni-ControlNet~\pcite{zhao2023uni} to enable musical controls that are \emph{composable} (i.e., can generate music corresponding to any subset of controls) and further allow creators to only
\textit{partially specify} each of the controls
both for convenience and to direct our model to musically \textit{improvise} in remaining time spans of the generation.
To overcome the aforementioned scarcity of precise, ground-truth control inputs, following~\citep{copet2023simple, donahue2023singsong},
we extract useful control signals directly from music during training.
We evaluate our method on two different categories of control signals: 
(1)~\emph{\extracted} control signals that come from example songs, which are similar to those seen during training, and 
(2)~\emph{\expected} control signals that we anticipate creators might want to use in a co-creation setting. 
Our experiments show that we can generate realistic music that accurately corresponds to control inputs in both settings. 
Moreover, we compare our approach against the melody control of the recently proposed MusicGen~\pcite{copet2023simple}, showing that our model is $49$\% more faithful to melody input, despite also controlling dynamics and rhythm, having $35$x fewer parameters, and being trained on $11$x less data.

Our contributions include:
\begin{itemize}[topsep=0pt,itemsep=1pt,leftmargin=*]
    \item A general framework for augmenting text-to-music models with 
    composable, 
    precise, \timevar{} musical controls.
    \item A method to enable one or more partially-specified \timevar{}  controls at inference. 
    \item Effective application of our framework to melody, dynamics, and rhythm control using music feature extraction algorithms together with conditional diffusion models.
    \item Demonstration that our model generalizes from extracted controls seen during training to ones we expect from creators. 
\end{itemize}

%% file: sections/background.tex
\section{Background: Diffusion and Image Generation}\label{sec:background}

\subsection{Diffusion Models}\label{subsec:diffusion}
We use denoising~diffusion~probabilistic~models~(DDPMs) \citep{sohl2015deep,ho2020denoising} as our underlying generative modeling approach for music audio.
DDPMs are a class of latent generative variable model. A DDPM generates data $\bm{x}^{(0)} \in \mathcal{X}$ from Gaussian noise $\bm{x}^{(M)} \in \mathcal{X}$ through a denoising Markov process that produces intermediate latents $\bm{x}^{(M-1)}, \bm{x}^{(M-2)},\dots,\bm{x}^{(1)} \in \mathcal{X}$, where $\mathcal{X}$ is the data space.
DDPMs can be formulated as the task of modeling the joint probability distribution of the desired output data $\bm{x}^{(0)}$ and all intermediate latent variables, i.e.,
\begin{equation}
    p_{\theta}(\bm{x}^{(0)},\dots,\bm{x}^{(M)}) \defeq p(\bm{x}^{(M)}) \prod_{m=1}^M p_{\theta}(\bm{x}^{(m-1)} | \bm{x}^{(m)}) \, , \label{eqn:joint-data-prob}
\end{equation}
where $\theta$ denotes the set of parameters to be learned, and $p(\bm{x}^{(M)}) \defeq \mathcal{N}(\bm{0}, \bm{I})$
is a fixed noise prior.

To create training examples, a forward diffusion process $q(\bm{x}^{(0)},\dots,\bm{x}^{(M)})$ is used to gradually corrupt clean data examples $\bm{x}^{(0)}$ via a Markov chain that iteratively adds noise:
\begin{equation}
\begin{aligned}
q(\bm{x}^{(0)},\dots,\bm{x}^{(M)}) &\defeq q(\bm{x}^{(0)}) \prod_{m=1}^M q(\bm{x}^{(m)} | \bm{x}^{(m-1)}) \,  \\
q(\bm{x}^{(m)} | \bm{x}^{(m-1)}) &\defeq \mathcal{N}(\sqrt{1 - \beta_m}\bm{x}^{(m-1)}, \beta_m \bm{I}),
\end{aligned}
\end{equation}
where $q(\bm{x}^{(0)})$ is the true data distribution, and $\beta_1,\dots,\beta_M$ are a sequence of parameters that define the noise level within the forward diffusion process, also known as the noise schedule.

By definition of $q(\bm{x}^{(m)} | \bm{x}^{(m-1)})$, it follows that the noised data $\bm{x}^{(m)}$ at any noise level $m \in \{1, \dots, M\}$ can be sampled in one step via:
\begin{equation}
    \bm{x}^{(m)} \defeq \sqrt{\bar{\alpha}_m}\bm{x}^{(0)}+ \sqrt{1-\bar{\alpha}_m}\bm{\epsilon} \, , \label{eqn:add-noise}
\end{equation}
where $\bar{\alpha}_m \defeq \prod_{m'=1}^{m} (1 - \beta_{m'})$,  $\bm{\epsilon} \sim \mathcal{N}(\bm{0}, \bm{I})$, and $M$ is the total number of noise levels or steps during training.
It was shown by Ho et al.~\cite{ho2020denoising} that we can optimize the variational lower bound~\citep{kingma2013auto} of the data likelihood, i.e., $p_\theta(\bm{x}^{(0)})$,
by training a function approximator, e.g., a neural network, ${f_\theta(\bm{x}^{(m)}, m): \mathcal{X} \times \mathbb{N} \rightarrow \mathcal{X}}$ to recover the noise $\bm{\epsilon}$ added via~\eref{eqn:add-noise}.
More specifically, $f_\theta(\bm{x}^{(m)}, m)$ can be trained by minimizing the mean squared error, i.e.,
\begin{equation}
    \mathbb{E}_{\bm{x}^{(0)}, \bm{\epsilon}, m}\Big[ \lVert \bm{\epsilon} - f_\theta(\bm{x}^{(m)}, m) \rVert^2_2 \Big] \, . \label{eqn:objective}
\end{equation}

With a trained $f_\theta$, we can transform random noise ${\bm{x}^{(M)} \sim \mathcal{N}(\bm{0}, \bm{I})}$ to a realistic data point $\bm{x}^{(0)}$ through $M$ denoising iterations.
To obtain high-quality generations, a large $M$ (e.g., $1000$) is typically used.
To reduce computational cost, denoising diffusion implicit models (DDIM)~\citep{song2020denoising} further
proposed an alternative formulation that allows running much fewer than $M$ sampling steps (e.g., $50\sim100$) at inference with minimal impact on generation quality.

\subsection{UNet Architecture for Image Diffusion Models}
Our approach to music generation is rooted in methodology developed primarily for generative modeling of images.
When applying diffusion modeling for image generation, the function $f_\theta$
is often a large UNet~\pcite{ho2020denoising,ronneberger2022convolutional}. 
The UNet architecture consists of two halves, an encoder and a decoder, that typically input and output image-like feature maps 
in the pixel space~\citep{saharia2022photorealistic} or some learned latent space~\pcite{rombach2022high}.
The encoder progressively downsamples the input to learn useful features at different resolution levels,
while the decoder, which has a mirroring architecture to the encoder and accepts features from corresponding encoder layers through skip connections, progressively upsamples the features to eventually get back to the input dimension.
For practical use,
diffusion-based image generation models are often text-conditioned,
which requires augmenting the network $f_\theta$ to accept a text description ${\bm{c}_\text{text}\in \mathcal{T}}$, where $\mathcal{T}$ is the set of all text descriptions.
This leads to the following function signature:
\begin{equation}
    f_\theta(\bm{x}^{(m)}, m, \bm{c}_\text{text}):  \mathcal{X} \times \mathbb{N} \times \mathcal{T} \rightarrow \mathcal{X}  \, ,
\end{equation}
which, via the process outlined in Sec.~\ref{subsec:diffusion}, models the desired probability distribution $p_\theta(\bm{x}^{(0)} \, | \, \bm{c}_\text{text})$. 
The text condition $\bm{c}_\text{text}$ is typically a sequence of embeddings from a large language model (LLM) or one or more embeddings from a learned embedding layer for class-conditional control. In either case, the conditioning signals $m$ (i.e., the diffusion time step) and $\bm{c}_\text{text}$ are usually incorporated in the UNet hidden layers via additive sinusoidal embeddings~\pcite{ho2020denoising} and/or cross-attention~\pcite{rombach2022high}.

\subsection{Classifier-free Guidance}
To improve the flexibility of text conditioning, classifier-free guidance (CFG) is commonly employed. CFG is used to simultaneously learn a conditional and unconditional generative model together and trade-off conditioning strength, mode coverage, and sample quality~\pcite{ho2022classifier}. Practically speaking, during training CFG is achieved by randomly setting conditioning information to a special null value $\bm{c}_\emptyset$ for a fraction of the time during training. Then during inference, an image is generated using conditional control inputs, unconditional control inputs, or a linear combination of both. In most cases, a forward pass of $f_\theta(\bm{x}^{(m)}, m, \bm{c}_\text{text})$ and $f_\theta(\bm{x}^{(m)}, m, \bm{c}_\emptyset)$ per sampling step are needed and subsequent weighted averaging. 

\subsection{Adding Pixel-level Controls to Image Diffusion Models}\label{subsec:adding-img-ctrl}
ControlNet~\pcite{zhang2023adding} proposed an effective method to add \textit{pixel-level} (i.e., \textit{spatial}) controls to large-scale pretrained text-to-image diffusion models.
Let the diffusion model input/output space be images, i.e., $\mathcal{X} \defeq \mathbb{R}^{W \times H \times D}$, where $W, H, D$ are respectively the width, height, and depth (for RGB images, $D = 3$) of an image,
we denote the set of $N$ pixel-level controls as:
\begin{equation}
\bm{C} \defeq \{\bm{c}^{(n)} \in \mathbb{R}^{W \times H \times D_n}\}_{n=1}^{N} \, ,\label{eqn:img-cond}  
\end{equation}
where $D_n$ is the depth specific to each $\bm{c}^{(n)}$.
For each condition signal $\bm{c}^{(n)}$,
every pixel $\bm{c}^{(n)}_{i, j} \in \mathbb{R}^{D_n}$, where $i \in \{1,\dots, W\}$ and $j \in \{1,\dots,H\}$, asserts an attribute on the corresponding pixel $\bm{x}^{(0)}_{i, j}$ in the output image. For example, ``$\bm{x}^{(0)}_{i, j}$ is (not) part of an edge'' or ``the perceptual depth of $\bm{x}^{(0)}_{i, j}$''. 
Naturally, the function to be learned, $f_\theta$, should be revised again as:
\begin{equation}
    f_\theta(\bm{x}^{(m)}, m, \bm{c}_\text{text}, \bm{C}):  \mathcal{X} \times \mathbb{N} \times \mathcal{T} \times \mathcal{C} \rightarrow \mathcal{X}  \, , \label{eqn:ctrlnet-func}
\end{equation}
where $\mathcal{C}$ denotes the set of all possible sets of control signals.
The updated $f_\theta$ hence implicity models $p_\theta(\bm{x}^{(0)} \, | \, \bm{c}_\text{text}, \bm{C})$.

To promote training data efficiency, 
ControlNet instantiates $f_\theta(\bm{x}^{(m)}, m, \bm{c}_\text{text}, \bm{C})$ by
reusing the pretrained (and frozen) text-conditioned UNet,
and clones its encoder half to form an adaptor branch to incorporate pixel-level control through finetuning.
To gracefully bring in the information from pixel-level control, it enters the adaptor branch through a convolution layer that is initialized to zeros (i.e., a \textit{zero convolution} layer).
Outputs from layers of the adaptor branch are then fed back to the corresponding layers of the frozen pretrained decoder, also through zero convolution layers, to influence the final output. 
Uni-ControlNet~\citep{zhao2023uni} then augmented the adaptor branch such that one model can be finetuned to accept multiple pixel-level controls via a single adaptor branch  without the need to specify all controls at once whereas ControlNet requires separate adaptor branches per control.

%% file: sections/method.tex
\section{Music ControlNet}
Our Music ControlNet framework builds on the methodology of text-to-image generation with pixel-level controls, i.e., ControlNet~\citep{zhang2023adding} and Uni-ControlNet~\citep{zhao2023uni}, and extends it for text-to-audio generation with time-varying controls.
We formulate our controllable audio generation task, explain the links and differences to ControlNet, and detail our essential model architecture and training modifications below.

\subsection{Problem Formulation}
Our overall goal is to learn a conditional generative model 
$p(\bm{w} \,|\, \bm{c}_\text{text}, \bm{C})$
over audio waveforms $\wave$, given a global (i.e., time-independent) text control $\bm{c}_\text{text}$, and a set of \timevar{} controls $\bm{C}$. Due to our dataset, we limit $\bm{c}_\text{text}$ to \textit{musical genre} and \textit{moods} tags.
Waveforms $\wave$ are vectors in $\mathbb{R}^{\mathrm{T}\mathrm{f_s}}$, where $\mathrm{T}$ is the length of audio in seconds and $\mathrm{f_s}$ is the sampling rate (i.e., number of samples per second).
As $\mathrm{f_s}$ is large (typically between 16~kHz and 48~kHz),
it is empirically difficult to directly model $p(\bm{w} \,|\, \cdot)$.
Hence, we adopt a common hierarchical approach of using spectrograms as an intermediary. 
A \textit{spectrogram} $\spec \in \mathbb{R}^{\mathrm{T}\mathrm{f_k} \times B \times D}$ is an image-like representation for audio signals, obtained through Fourier Transform on $\wave$,
where $\mathrm{f_k}$ is the frame rate (usually 50$\sim$100 per second), $B$ is the number of frequency bins, and $D=1$ for mono-channel audio.
With $\spec$ as the intermediary, we instead model the joint distribution 
${p(\wave, \spec \,|\, \bm{c}_\text{text}, \bm{C})}$, which can be factorized as:
\begin{align}
    p(\wave, \spec \,|\, \bm{c}_\text{text}, \bm{C}) =& \; p(\wave \,|\, \spec, \bm{c}_\text{text}, \bm{C}) \cdot p(\spec \,|\, \bm{c}_\text{text}, \bm{C}) \\
    \defeq& \; p_\phi(\wave \, | \, \spec) \cdot p_\theta( \spec \,|\, \bm{c}_\text{text}, \bm{C}) \, ,
\end{align}
where $\phi$ and $\theta$ are sets of parameters to be learned.
Note that this factorization assumes conditional independence between waveform $\wave$ and all control signals $\bm{c}_\text{text}$ and $\bm{C}$ given spectrogram $\bm{s}$,
which is reasonable if the time-varying controls in $\bm{C}$ vary at a rate no faster than $\mathrm{f_k}$ by nature.

In our work,
\textbf{we focus on modeling spectrograms given controls}, i.e., $p_\theta( \spec \,|\, \bm{c}_\text{text}, \bm{C})$, and directly apply the DiffWave \textit{vocoder}~\citep{kong2020diffwave} to model $p_\phi(\wave \, | \, \spec)$.
Following the text-to-image ControlNet~\citep{zhang2023adding} model, we leverage diffusion models~\citep{ho2020denoising} to learn $p_\theta( \spec \,|\, \bm{c}_\text{text}, \bm{C})$.
If we set the input space ${\mathcal{X} \defeq \mathbb{R}^{\mathrm{T}\mathrm{f_k} \times B \times D}}$, and the desired output $\bm{x}^{(0)} \defeq \spec$, we can instantiate a neural network $f_\theta$ having an identical function signature to Eqn.~(\ref{eqn:ctrlnet-func}).
However, we observe two key differences between pixel-level controls for images and time-varying controls for audio/music.

First, the first two dimensions in a spectrogram $\spec$ have different semantic meanings, one being \textit{time} and the other being \textit{frequency}, as opposed to both being spatial in an image.
Second, the time-varying controls useful to creators are closely coupled with \textit{time}, but could have a much more relaxed relationship with \textit{frequency} such that the second dimension of~\eref{eqn:img-cond} cannot be restricted to $B$.
For example,
an intuitive control over `musical dynamics' may involve defining volume over time, not over frequency.
A high dynamics value for one frame can mean a number of different profiles over the $B$ frequency bins for the corresponding spectrogram frame, e.g., a powerful bass playing a single pitch, or a rich harmony of multiple pitches, which the model has freedom to decide.
Therefore, we relax the definition for the set of $N$ control signals to become:
\begin{equation}
    \bm{C} \defeq \{\bm{c}^{(n)} \in \mathbb{R}^{\mathrm{T}\mathrm{f_k} \times B_n \times D_n}\}_{n=1}^{N} \, , \label{eqn:music-conds}
\end{equation}
where $B_n$
is the number of classes 
 specific to each control $\bm{c}^{(n)}$, which is not bound to $B$.
With this updated definition, the correspondence between control signals $\bm{C}$ and the output spectrogram $\bm{x}$ naturally becomes \textit{frame-wise}.
For example, suppose $\bm{c}^{(n)}$ represents dynamics control, a frame for the control ${\bm{c}^{(n)}_{t} \in \mathbb{R}^{1 \times 1}}$, where $t \in \{1,\dots,\mathrm{T}\mathrm{f_k}\}$, then describes ``the musical dynamics (intensity) of the spectrogram frame~$\spec_t$''.

Finally, we consider time-varying controls $\bm{c}^{(n)}$ that can be directly extracted from spectrograms.
Given that spectrograms are also computed directly from waveforms,
only pairs of ${(\wave, \bm{c}_\text{text})}$ are necessary for training, causing no extra annotation overhead.
Nevertheless, we note that our formulation supports manually annotated time-varying controls as well.

\subsection{Adding Time-varying Controls to Diffusion Models}\label{subsec:ctrl-inject}
We propose a strategy to learn the mapping between input controls and the frequency axis of output spectrograms, marking an update from the ControlNet~\citep{zhang2023adding} method for image modeling.
As mentioned in~\sref{subsec:adding-img-ctrl}, ControlNet clones the encoder half of the pretrained UNet for text-to-image generation as the \textit{adaptor} branch, which uses newly attached zero convolution layers to enable pixel-level control.
Let $\Tilde{f}^{(l)}(\bm{x}^{(m, l-1)}, m, \bm{c}_\text{text}, \bm{C})$ denote the $l^\text{th}$ block of the adaptor branch,
where $m$ is the diffusion time step, $\bm{x}^{(m, l-1)}$ contains the features of the noised image after $l - 1$ blocks, and $\bm{c}_\text{text}$, $\bm{C}$ are the text and pixel-level controls respectively.
Considering the case $\bm{C} \defeq \{\bm{c}^{(1)}\}$ which is consistent with past work~\citep{zhang2023adding}, the pixel-level control is incorporated via:
\begin{equation}
    \begin{aligned}
    &{\Tilde{f}}^{(l)}(\bm{x}^{(m, l-1)}, m, \bm{c}_\text{text}, \bm{C}) \defeq\\ 
    &\mathcal{Z}_\mathrm{out}(f^{(l)}(\bm{x}^{(m, l-1)}+ \mathcal{Z}_\mathrm{in}(\bm{c}^{(1)}), m, \bm{c}_\text{text})) \, , \label{eqn:ctrlnet-inject}
    \end{aligned}
\end{equation}
where $\mathcal{Z}_\mathrm{in}$ and $\mathcal{Z}_\mathrm{out}$ are the newly attached zero convolution layers, and $f^{(l)}$ is initialized from the $l^\text{th}$ encoder block of the pretrained text-conditioned UNet.

In Music ControlNet, we revamp the control process to be:
\begin{equation}
    \begin{aligned}
            &\Tilde{f}^{(l)}(\bm{x}^{(m, l-1)}, m, \bm{c}_\text{text}, \bm{C}) \defeq \\
            &\mathcal{Z}_\mathrm{out}(f^{(l)}(\bm{x}^{(m, l-1)} + \mathcal{Z}_\mathrm{in}(\mathcal{M}(\bm{c}^{(1)})), m, \bm{c}_\text{text})) \, , 
    \end{aligned}
\label{eqn:musctrl-inject}
\end{equation}
where $\mathcal{M}$ is an additional 1-hidden-layer MLP that transforms $B_1$, the number of classes for the control $\bm{c}^{(1)}$ following~\eref{eqn:music-conds}, to match the number of frequency bins $B$, and simultaneously learns the relationship between control classes and frequency bins.
In cases with multiple controls, i.e., $\bm{C} =\{\bm{c}^{(n)}\}_{n=1}^{N}$, each control is processed with its individual MLP, i.e., $\mathcal{M}^{(n)}$, and then concatenated along the depth dimension, i.e., $D_n$, before entering the shared zero-convolution layer $\mathcal{Z}_\mathrm{in}$.

\begin{figure}
    \centering
    \includegraphics[width=\columnwidth, trim={2.5mm, 0, 2.5mm, 0}, clip]{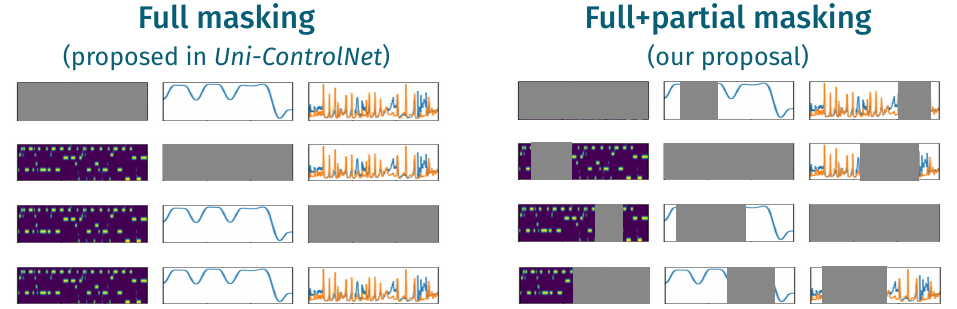}
    \caption{Two masking schemes we randomly choose to apply during training that allow creators to input \textit{any subset} of the time-varying controls, \textit{fully} or \textit{partially} specified in time, at inference.
    Each row indicates a unique masking instantiation over the set of control signals $\bm{C} \defeq \{\bm{c}^{(n)}\}_{n=1}^{N}$ ($N = 3$ as illustrated here).
    Masked control signals are colored in {\color{gray} gray}.}
    \label{fig:masking}
\end{figure}

\subsection{Masking Strategy to Enable Partially-specified 
Controls}\label{subsec:masking}
To give creators the freedom to input any subset of the $N$ controls, Uni-ControlNet~\citep{zhao2023uni} proposed a CFG-like training strategy to drop out each of the control signals $\bm{c}^{(n)}$ randomly during training. We follow the same strategy and further assign a higher probability to keep or drop all controls~\citep{zhao2023uni} as we found that this leads to perceptually better generations.
In more detail, we let the index set of control signals be $\mathcal{I} = \{1,\dots,N\}$. 
At each training step, we then select a subset $\mathcal{I}' \subseteq \mathcal{I}$ that will be set to zero or dropped. 
We then directly apply the index subset to the control signals via:
\begin{equation}
    \bm{c}^{(n)} := 
    \begin{cases}
    \bm{0}_{\mathrm{T}\mathrm{f_k} \times B_n \times D_n} \; &\forall n \in \mathcal{I}' \, \\
    \bm{c}^{(n)} \; &\forall n \in \mathcal{I} \setminus \mathcal{I}' \, .
    \end{cases}
    \label{eqn:unictrl-drop}
\end{equation}
Doing so induces $f_\theta(\bm{x}^{(m)}, m, \bm{c}_\text{text}, \bm{C})$ to learn the correspondence between any subset of the $N$ control signals and the output spectrogram.

In Music ControlNet, we further desire a model that allows the given subset of controls to be \textit{partially-specified in time}.
Therefore, we devise a new scheme that partially masks the active controls (i.e., those indexed by $\mathcal{I} \setminus \mathcal{I}'$)
Specifically, we randomly sample a pair $(t_{n, a}, t_{n, b}) \in \{1,\dots,\mathrm{T}\mathrm{f_k}\}^2$, where $t_{n, a} < t_{n, b}$, for each of the active controls, and mask them as:
\begin{equation}
    \bm{c}^{(n)}_{t} := 
    \begin{cases}
    \bm{0}_{B_n \times D_n} \; &\mathrm{if} \; t \in [t_{n, a},t_{n, b}] \, \\
    \bm{c}^{(n)}_{t} \; &\mathrm{otherwise}
    \end{cases}  \;\;\; \forall n \in \mathcal{I} \setminus \mathcal{I}'
    \label{eqn:linked-drop}
\end{equation}
Fig.~\ref{fig:masking} displays example instantiations of the two masking schemes detailed above.
At each training step, after selecting $\mathcal{I}'$ (i.e., determining the dropped controls) we choose one of the two masking schemes uniformly at random, and then sample the timestamp pairs (i.e., $(t_{n, a}, t_{n, b})$'s) when needed. In this way, we further employ a CFG-like training strategy to enable partially-specified controls in a unified manner.

\subsection{Musical Control Signals}\label{subsec:music-ctrl}
Here, we outline the time-varying controls, i.e., $\bm{C}$, that we combine with our proposed framework to build a music generation model that is useful for creators.
We propose three control signals, namely, \textit{melody}, \textit{dynamics}, and \textit{rhythm}, that are musically complementary to each other.
These controls are designed such that they can be directly extracted from the target spectrogram, i.e, $\spec \in \mathbb{R}^{\mathrm{T}\mathrm{f_k} \times B \times D}$, requiring no human annotation at all,
and allow music creators to easily \textit{create} their control signals at inference time to compose their music from scratch,
in addition to \textit{remixing}, i.e.,  combining musical elements from different sources, using controls \textit{extracted} from existing music.
Below, we briefly introduce how our control signals are obtained, what connections they have to music, and how creators can form \expected{} controls at inference time.
Readers may refer to Fig.~\ref{fig:single-ctrls} for how the control signals may be visually presented to creators.

\begin{itemize}[leftmargin=*, topsep=3pt, itemsep=3pt]
    
    \item \textbf{Melody ($\bm{c}_{\mathrm{mel}} \in \mathbb{R}^{\mathrm{T}\mathrm{f_k} \times 12 \times 1}$): \;}
Following~\pcite{copet2023simple}, we adopt a variation of the chromagram~\pcite{muller2015fundamentals} to encode the most prominent musical tone over time. 
To do so, we compute a linear spectrogram and then rearrange the energy across the $B$ frequency bins into $12$ pitch classes (or semitones, i.e., C, C-sharp, \dots, B-flat, B) in a frame-wise manner, i.e., independently for each $t \in \{1, \dots, \mathrm{T}\mathrm{f_k}\}$, via the Librosa Chroma function~\cite{librosa}.
To form a better proxy for melody from the raw chromagram, only the most prominent pitch class is preserved by applying an $\mathrm{argmax}$ operation to make the chromagram frame-wise one-hot.
Additionally, we apply a Biquadratic high-pass filter~\cite{yang2021torchaudio} with a cut-off at Middle C, or 261.2 Hz) before chromagram computation to avoid bass dominance, i.e., the resulting one-hot chromagram encodes the bass notes, rather than the desired melody notes.
At test time, the melody control can be created by recording a simple melody, or simply drawing the pitch contour.
A desirable model should be able to turn the simple created melody control into rich, high-quality multitrack music.
    
    \item \textbf{{Dynamics ($\bm{c}_{\mathrm{dyn}} \in \mathbb{R}^{\mathrm{T}\mathrm{f_k} \times 1 \times 1}$)}: \;}
The dynamics control is obtained by summing the energy across frequency bins per time frame of a linear spectrogram, and mapping the resulting values to the decibel (dB) scale, which is closely linked to loudness perceived by humans~\cite{librosa}.
To mitigate rapid fluctuations of the raw dynamic values due to note or percussion onsets, and also to bring our dynamics control closer to the perceived musical intensity, we apply a smoothing filter with one second context window over the frame-wise values (i.e., a Savitzky-Golay filter~\cite{scipy}). The dynamics control not only characterizes the loudness of notes, but also is strongly correlated with important musical intensity-related attributes like instrumentation, harmonic texture, and rhythmic density thanks to the natural correlation between loudness and the aforementioned attributes in human-composed music.
During inference, creators can simply draw a line/curve of how they want the musical intensity to vary over time as the created dynamics control.
    
    \item \textbf{Rhythm ($\bm{c}_{\mathrm{rhy}} \in \mathbb{R}^{\mathrm{T}\mathrm{f_k} \times 2 \times 1}$): \;}
For rhythm control, we employ an in-house implementation of an RNN-based beat detector~\pcite{bock2016madmom} that is trained on a different internal dataset to predict whether a frame is situated on a beat, a downbeat, or neither. We then use the frame-wise \textit{beat} and \textit{downbeat} probabilities for control, resulting in $2$ classes per frame.
The advantages of our time-varying beat/downbeat control over just inputting a global tempo (i.e, beats per minute) are:
(i) it allows creators to precisely synchronize beats/downbeats with, for example, video scene cuts or other moments of interest in the content to be paired with generated music.
(ii) it encodes some nuanced information of rhythmic feeling, e.g., whether the music sounds more harmonic or rhythmic, and whether the rhythmic pattern is clear/simple, or complex, on which experienced music creators may want to influence in the generative process.
At inference, the rhythm control can be created by time-stretching the beat/downbeat probability curves extracted from existing songs to match the desired tempo.
Also, creators can obtain precise beat/downbeat timestamps by feeding the beat/downbeat curves to a Hidden Markov Model (HMM) based post-filter~\pcite{krebs2015efficient, bock2016joint}, and use the timestamps to shift the curves along the time axis for synchronization purposes mentioned above. We also tried to manually draw spiked curves as the created rhythm control, but the performance of this was worse than our final hand-drawn (i.e., created) dynamics control.
\end{itemize}

%% file: sections/experimental_design.tex
\begin{table*}
\footnotesize
\caption{Performance of single vs.~multiple time-varying controls using controls \textit{\extracted{}} from our in-domain test set. Time-varying controllability (i.e., Melody, Dynamics, Rhythm metrics) is notably higher when the corresponding control is passed to the model (\cmark) as opposed to being excluded (\xmark). 
\textbf{Bold} values indicates enforced controls. Higher is better except for FAD.
}\label{tab:single-vs-multi}
\vspace{-2mm}
\renewcommand\arraystretch{1.1}
\centering
\begin{tabular}{l | c c c  | c | c c | c c | c c }
\toprule
& \multicolumn{3}{c|}{\textbf{Control signals}} & \textbf{Melody} acc (\%) & \multicolumn{2}{c|}{\textbf{Dynamics} corr ($r$, in \%)} & \multicolumn{2}{c|}{\textbf{Rhythm} F1 (\%)}  & 
 \textbf{CLAP} & \textbf{FAD} $\downarrow$ \\
 & $\mathrm{mel}$ & $\mathrm{dyn}$ & $\mathrm{rhy}$ &  & Micro & Macro & Beat & Downbeat &  \\
\midrule
\textbf{Global style only} & \xmark & \xmark & \xmark  &  \;\,8.5 & \hspace{-1mm}$-$0.7 & \;\,0.7 & 27.8 & \;\,7.8 & 0.28 & 1.51 \\ \hline
\multirow{3}{*}{\textbf{Single controls}} & \cmark & \xmark & \xmark  &  \textbf{58.3} & \;\,4.4 & \;\,3.1 & 40.2 & 12.1 & 0.28 & 1.34 \\
& \xmark & \cmark & \xmark  &  \;\,8.6 & \textbf{88.8} & \textbf{63.6} & 36.7 & 16.1 & 0.26 & 1.50 \\ 
& \xmark & \xmark & \cmark  &  \;\,8.6 & 25.8 & 34.6 & \textbf{69.2} & \textbf{35.4} & 0.27 & 1.17 \\ \hline
\multirow{4}{*}{\textbf{Multi controls}} & \cmark & \cmark & \xmark  &  \textbf{57.7} & \textbf{89.7} & \textbf{64.8} & 47.4 & 21.8 & 0.26 & 1.38 \\
& \cmark & \xmark & \cmark  & \textbf{59.1} & 31.6 & 36.3 & \textbf{70.0} & \textbf{38.7} & 0.26 & 1.16 \\
& \xmark & \cmark & \cmark  &  \;\,8.7 & \textbf{89.6} & \textbf{60.9} & \textbf{72.1} & \textbf{39.9} & 0.26 & 1.12 \\
& \cmark & \cmark & \cmark  & \textbf{58.7} & \textbf{90.8} & \textbf{64.0} & \textbf{70.8} & \textbf{40.8} & 0.25 & 1.14 \\
\bottomrule
\end{tabular}
\end{table*}

\section{Experimental Setup}

\subsection{Datasets}\label{subsec:dataset}
We train our models on a dataset of  ${\approx}1800$ hours of licensed instrumental music with genre and mood tags.
Our dataset does not have free-form text description,
so we use class-conditional text control of global musical style, as done in JukeBox~\citep{dhariwal2020jukebox}. 
For evaluation, we use data from four sources:
(i)~an \textbf{in-domain test set} with 2K songs held out from our dataset, 
(ii)~the \textbf{MusicCaps dataset}~\pcite{agostinelli2023musiclm} with around 5K 10-second clips associated with free-form text description
(iii)~the \textbf{MusicCaps+ChatGPT dataset} where we use ChatGPT~\pcite{chatgpt} to convert the free-form text in MusicCaps to mood and genre tags that match our dataset via the prompt ``\textit{For the lines of text below, convert each to one of the following [genres or moods] and only output the [genre or mood] per line (no bullets):  [MusicCaps description]}'', and 
(iv)~a \textbf{Created Controls dataset} of control signals that music creators can realistically give via manual annotation or similar. 

\subsection{Created Controls Dataset Details}\label{subsec:creator_dataset}
For our Created Controls dataset, we created example melodies, dynamics annotations, and rhythm presets that we envision creators would use during music co-creation via:
\begin{itemize}[leftmargin=*, topsep=2pt, itemsep=2pt]
    \item \textbf{Melody:} We record our piano play of 10 well-known classical public domain music melodies (30 seconds long each) composed by Bach, Vivaldi, Mozart, Beethoven, Schubert, Mendelssohn, Bizet, and Tchaikovsky, and crop two 6-second chunks, minimizing repeated musical content as possible, resulting in a 20-example melody controls.
    \item \textbf{Dynamics:}  To simulate a creator-drawn dynamics curves, we draw out 6-second long dynamics curves as $\{\mathrm{Linear, Tanh, Cosine}\}$ functions, either vertically flipped or not, with scaled dynamics ranges of $\{\pm6, \pm9, \pm12, \pm15\}$ decibels from the mean value of all training examples.
    This leads to 3$\times$2$\times$4$\,=\,$24 created dynamics controls.
    \item \textbf{Rhythm:} We create ``rhythm presets'' via selecting four songs from our \textbf{in-domain test set} with different rhythmic strengths and feelings, extract their rhythm control signals, and time-stretch them using PyTorch interpolation with factors $\{0.8, 0.9, 1.0, 1.1, 1.2\}$ to create 20 rhythm controls.
\end{itemize}
Each set of created controls is then cross-producted with 10 genres $\times$ 10 moods to form the final dataset of 2.0K, 2.4K and 2.0K samples. Our created controls are distinct from controls that are directly extracted from mixture data during training. 

\subsection{Model, Training, and Inference Specifics}
For our spectrogram generation model $p_\theta( \spec \,|\, \bm{c}_\text{text}, \bm{C})$,
we use a convolutional UNet~\citep{ronneberger2022convolutional} with 5 2D-convolution ResNet~\citep{he2016deep} blocks with $[64, 64, 128, 128, 256]$ feature channels per block with a stride of 2 in between downsampling blocks. The UNet inputs Mel-scaled~\pcite{stevens1937scale} spectrograms clipped to a dynamic range of 160 dB and scaled to $[-1, 1]$  computed from 22.05 kHz audio with a hop size of 256 (i.e., frame rate $\mathrm{f_k} \approx 86$ Hz), a window size of 2048, and 160 Mel bins. For our genre and mood global style control $\bm{c}_\text{text}$, we use learnable class-conditional embeddings with dimension of 256 that are injected into the inner two ResNet blocks of the U-Net via cross-attention. We use a cosine noise schedule with 1000 diffusion steps $m$ that are injected via sinusoidal embeddings with a learnable linear transformation summed directly with U-Net features in each block. To approximately match the output dimensions of ControlNet (512$\times$512$\times$3), we set our output time dimension to 512 or $\approx$6 seconds, yielding a 512$\times$160$\times$1 output dimension. We use an L1 training objective between predicted and actual added noise, an Adam optimizer with learning rate to $10^{-5}$ with linear warm-up and cosine decay. Due to limited data and efficiency considerations, we instantiate a relatively small model of 41 million parameters and pretrain with distributed data parallel for 5 days on 32 A100 GPUs with a batch size of 24 per GPU. 

Given our pretrained global style control model, we finetune on time-varying melody, dynamics, and rhythm controls controls. The time-varying controls enter the pretrained U-Net via an adaptor branch as discussed above. We use the same loss and optimizer used for pretraining and finetune until convergence for 3 days with 8 A100 GPUs.
At inference, we use 100-step DDIM~\pcite{song2020denoising} sampling, and CFG~\cite{ho2022classifier} on global style control with a scale of 4 on the global style control only.

For our spectrogram-to-audio vocoder $p_\phi(\wave \, | \, \spec)$, we train a diffusion-based DiffWave~\citep{kong2020diffwave} vocoder. We leverage an open-source package~\pcite{lmntdiffwave}, and use our main training dataset, an Adam optimizer with learning rate of $10^{-5}$, noise prediction L1 loss, a 50-step linear noise schedule,
hopsize of 256 samples, sampling rate of 22050Hz, batch size of 50 per GPU and train on 8 GPUs for 10 days. For inference, we adopt DDIM-like fast sampling~\citep{kong2020diffwave} with six steps.


\subsection{Evaluation Metrics}

We use the following metrics to evaluate time-varying controllability, adherence to global text (i.e., mood \& genre tags) control, and overall audio realism.
\begin{itemize}[topsep=2pt, itemsep=2pt, leftmargin=*]
    \item \textbf{Melody accuracy} examines whether the frame-wise pitch classes (C, C\#,\dots, B; 12 in total) match between the input melody control and that extracted from the generation.
    \item \textbf{Dynamics correlation} is the Pearson's correlation between the frame-wise input dynamics values to the values computed from the generation.
    We compute two types of correlation, which we call \textit{micro} and \textit{macro} correlation respectively.
    \textit{Micro} computes $r$'s separately for each generation, while \textit{macro} collects input/generation dynamics values from all generations, and then computes a single $r$. 
    The \textit{micro} correlation examines whether relative dynamics control values \textit{within a generation} is respected, while the \textit{macro} one checks the same property \textit{across many generations}.
    \item \textbf{Rhythm F1} 
    follows the standard evaluation methodology for beat/downbeat detection~\pcite{davies2009evaluation, raffel2014mir_eval}.
    It quantifies the alignment between the beat/downbeat timestamps estimated from the input rhythm control, and those from the generation.
    The timestamps are estimated by applying an HMM post-filter~\pcite{krebs2015efficient} on the frame-wise (down)beat probabilities (i.e., the rhythm control signal).
    Following~\citep{raffel2014mir_eval}, a pair of input and generated (down)beat timestamps are considered aligned if they differ by $<70$ milliseconds.
    \item \textbf{CLAP} score~\pcite{wu2023large, htsatke2022} evaluate text control adherence via computing the pair-wise cosine similarity of text and audio embeddings extracted from CLAP.
    CLAP is a dual-encoder foundation model where the encoders respectively receive a text input and an audio input.
    The text and audio embedding spaces are learned via a contrastive objective~\citep{oord2018representation}.
    To obtain the embeddings for evaluation, we feed the generated audio to the CLAP audio encoder, and
    set the CLAP text encoder input to ``\textit{An audio of [mood] [genre] music}'' to accommodate our tag-based control on global musical style. 
    \item \textbf{FAD} is the Fr\'echet distance between the distribution of embeddings from a set of reference audios and that from generated audios~\pcite{kilgour2018fr}. 
    It measures `\textit{how realistic the set of generated audios are}', taking both quality and diversity into account.
    To ensure comparable FAD scores, we utilize the Google Research FAD package~\pcite{kilgour2018fr}, which employs a VGGish~\pcite{hershey2017cnn} model trained on audio classification~\pcite{gemmeke2017audio} to extract embeddings from audios.
    Unless otherwise specified, the reference audios for FAD are our in-domain test dataset.
\end{itemize}

%% file: sections/results.tex

\begin{table*}
\footnotesize
\caption{Evaluation on controls \textit{\expected{}} by creators that are more simple than the \extracted{} controls seen by our model during training.
Using \expected{} controls leads to better time-varying controllability, i.e., Melody, Dynamics, Rhythm metrics.
}\label{tab:out-results}
\vspace{-2mm}
\renewcommand\arraystretch{1.1}
\centering
\begin{tabular}{l l | c | c c | c c | c c }
\toprule
Control & Control source & \textbf{Melody} acc (\%) & \multicolumn{2}{c|}{\textbf{Dynamics} corr ($r$, in \%)} & \multicolumn{2}{c|}{\textbf{Rhythm} F1 (\%)}  & 
 \textbf{CLAP} & \textbf{FAD} $\downarrow$ \\
 &  &  & Micro & Macro & Beat & Downbeat &  \\
\midrule
\multirow{2}{*}{Melody} &  \textbf{Extracted} &  58.3 & --- & --- & --- & ---& 0.28 & 1.34 \\
& \textbf{Created}  &  78.2 & --- & --- & --- & --- & 0.27 & 1.81 \\ \hline
\multirow{2}{*}{Dynamics} &  \textbf{Extracted} &  --- & 88.8 & 63.6 & --- & --- & 0.26 & 1.50 \\
& \textbf{Created}  &  --- & 98.5 & 93.2 & --- & --- & 0.26 & 2.18 \\ \hline
\multirow{2}{*}{Rhythm} &  \textbf{Extracted} & --- & --- &--- & 69.2 & 35.4 & 0.26 & 1.17 \\
& \textbf{Created}  &  --- & --- &--- & 88.6 & 45.2 & 0.26 & 2.93 \\ 
\bottomrule
\end{tabular}
\end{table*}

\begin{table*}
\footnotesize
\caption{Evaluation on controls \textit{partially specified} in time, which lift the requirement for creators to always input full controls.
Time-varying contrallability (Melody, Dynamics, Rhythm metrics) only degrades mildly with partially-specified controls.
}\label{tab:dontcare-results}
\vspace{-2mm}
\renewcommand\arraystretch{1.1}
\centering
\begin{tabular}{l l | c | c c | c c | c c }
\toprule
Control & Control source \& span & \textbf{Melody} acc (\%) & \multicolumn{2}{c|}{\textbf{Dynamics} corr ($r$, in \%)} & \multicolumn{2}{c|}{\textbf{Rhythm} F1 (\%)}  & 
 \textbf{CLAP} & \textbf{FAD} $\downarrow$ \\
 &  &  & Micro & Macro & Beat & Downbeat &  \\
\midrule
\multirow{2}{*}{Melody} &  \textbf{\Expected, full}  &  78.2 & --- & --- & --- & --- & 0.27 & 1.81 \\
& \textbf{\Expected, partial}   &  74.3 & --- & --- & --- & --- & 0.27 & 1.66 \\ \hline
\multirow{2}{*}{Dynamics} &  \textbf{\Expected, full} & --- &  98.5 & 93.2 & --- & --- & 0.26 & 2.18 \\
& \textbf{\Expected, partial} & --- &  88.6 & 89.0 & --- & --- & 0.27 & 1.52 \\ \hline
\multirow{2}{*}{Rhythm} &  \textbf{\Expected, full} &  --- & --- &--- & 88.6 & 45.2 & 0.26 & 2.93 \\
& \textbf{\Expected, partial} &  --- & --- &--- & 80.1 & 34.8 & 0.26 & 2.60 \\ 
\bottomrule
\end{tabular}
\end{table*}

\section{Evaluation and Discussion}\label{sec:results}
We conduct a comprehensive evaluation of our proposed Music ControlNet framework.
Specifically, we perform quantitative studies of
(i)~single vs.~multiple time-varying \extracted{} controls,
(ii)~\extracted{} controls vs.~\expected{} controls,
(iii)~fully vs.~partially-specified \expected{} controls, 
(iv)~extrapolating generation duration beyond the training duration (i.e., 6 seconds),
and (v)~benchmarking with the 1.5 billion-parameter MusicGen model with melody control. 
In all experiments, a single fine-tuned model is used with different inference configurations.
The duration of generation is 12 or 24 seconds in experiment (iv), 10 seconds in experiment (v) so we can be consistent with MusicCaps~\citep{agostinelli2023musiclm} benchmark, and 6 seconds in all other experiments. We leverage the fully convolutional nature of our UNet backbone to generate music that is longer than what is seen during training.
We conclude with an in-depth qualitative analysis of \expected{} generation examples.

\subsection{Single \& Multiple Extracted Controls}
We evaluate generation performance by applying different combinations of controls at inference time, using single or multiple control signals \textit{\extracted{}} from our \textbf{in-domain test set}.
The results are shown in Table~\ref{tab:single-vs-multi}.
First, we compare generations using global style only (i.e., genre and mood tags) and those with single \timevar{} controls (rows 2$\sim$4).
When the corresponding controls are enforced, we observe much higher melody accuracy, dynamics correlations, and rhythm F1s,
which indicate that
our proposed control injection mechanism (see Sec.~\ref{subsec:ctrl-inject}) affords effective time-varying controllability.
Interestingly, we find in rows 3 and 4 that the dynamics and rhythm metrics are higher compared to using global style control only (1${^\text{st}}$ row) even when the corresponding controls are excluded.
We hypothesize that this is due to that our rhythm and dynamics controls have natural correlation. 

Second, focusing on generations with multiple controls (last 4 rows in Table~\ref{tab:single-vs-multi}), we find the time-varying controllability metrics to remain largely the same compared to single control scenarios.
This shows that our model learns to simultaneously respond to multiple controls well despite the 
added complexity.
However, as more time-varying controls are enforced, text control adherence (CLAP score) degrades mildly, while overall audio realisticness (FAD) is not negatively impacted.

\subsection{From Extracted to Created Controls}\label{sec:generalization}
To empower creators to generate music with their own ideas, we evaluate the single-control generations using \textit{created} controls from our \textbf{Created Controls dataset}.
The comparison with \extracted{} controls are displayed in Table~\ref{tab:out-results}.
We notice several interesting insights.
First and perhaps unexpectedly, we find that across all three control signals, all time-varying controllability metrics actually improve when using \expected{} controls.
This demonstrates our model's generalizability to out-of-domain control inputs.

Second, we find that global style control adherence (CLAP score) is largely unaffected, while FAD appears to degrade. 
The degradation in FAD is multifaceted.
On the one hand, the \expected{} controls, naturally creates some music that is distributionally different from the in-domain test set.
Hence, we can not expect the desirable generations to score a low FAD.
On the other hand, perceptually, we do find the generations with \expected{} controls are more often less musically interesting.
We find this true particularly for \expected{} melody and dynamics controls, where the model may copy the melody with a single instrument on a constant background chord, or match dynamics using monotonous bass or sound effects.
However, in practice, we believe this is not an issue as creators can ask for a batch of generations and select the best one.


\subsection{From Fully- to Partially-specified Controls}
We evaluate generation quality using partially-specified, \expected{} control signals (made possible by the masking scheme in Sec.~\ref{subsec:masking}) and compare fully-specified \expected{} controls in Table~\ref{tab:dontcare-results}.
For partially-specified cases, for each sample, we specify the control for a random 1.0 to 4.5-second span out of the full 6-second duration.
The melody, dynamics, rhythm metrics  are computed only within the partially-specified spans,
while CLAP and FAD still take the full generated audio as input.
Overall, we find that partial control somewhat degrades time-varying controllability compared to the full \expected{} control scenarios, but it remains strong and mostly better than using full \extracted{} controls (cf.~rows marked by \textbf{Extracted} in Table~\ref{tab:out-results}).
Global style control adherence (CLAP) is unaffected. 
Overall quality (FAD) improves, suggesting that the less amount of controls induces the generations to match the training distribution better. We also found  that the coexistence of controlled and uncontrolled spans did not lead to pronounced incoherence issues.


\subsection{Extrapolating Duration of Generation}

\begin{table}[t]
\centering
\vspace{-2mm}
\footnotesize
\caption{Evaluation of generations of longer durations than that seen at training (i.e., 6 sec), using \textit{\expected{}} melodies. 
}\label{tab:extrap-results}
\vspace{-2mm}
\begin{tabularx}{7cm}{L c c c }
\toprule
Length & \textbf{Melody} acc(\%) & \textbf{CLAP} & \textbf{FAD} $\downarrow$   \\
\midrule
\textbf{6 sec}  & 78.2 & 0.27 & 1.81 \\
\textbf{12 sec} & 81.0 & 0.32 & 2.11 \\
\textbf{24 sec} & 82.8 & 0.33 & 2.54 \\
\bottomrule
\end{tabularx}
\end{table}

\begin{table*}[ht]
\footnotesize
\caption{Comparison to MusicGen~\pcite{copet2023simple} on the MusicCaps dataset~\pcite{agostinelli2023musiclm}.
Input melodies are either \textit{\extracted{}} from MusicCaps recordings or randomly selected from $20$ 
melodies from our \textit{created} controls dataset.
Our model exhibits more precise melody control, especially on created melodies, with comparable text control adherence when restricting MusicGen text prompts to our dataset's mood and genre tags (i.e., \textbf{CLAP}$_{\text{tag}}$).
Note that our model (41M parameters) is much smaller than MusicGen (1.5B parameters), and additionally
accepts multiple controls and partially-specified spans.
}\label{tab:vs-musicgen}
\vspace{-2mm}
\renewcommand\arraystretch{1.2}
\addtolength{\tabcolsep}{-0.36em}
\centering
\begin{tabular}{l l | c c c c c | c c c c c }
\toprule
Control & Model & \multicolumn{5}{c|}{\textbf{\Extracted} melody control} &  \multicolumn{5}{c}{\textbf{\Expected} melody control}\\
& & \textbf{Melody} & \textbf{CLAP}$_{\text{tag}}$ & \textbf{CLAP}$_{\text{text}}$ & \textbf{FAD}$_{\text{MCaps}}\downarrow$& \textbf{FAD}$_{\text{ours}}\downarrow$& \textbf{Melody} & \textbf{CLAP}$_{\text{tag}}$ & \textbf{CLAP}$_{\text{text}}$ & \textbf{FAD}$_{\text{MCaps}}\downarrow$ & \textbf{FAD}$_{\text{ours}}\downarrow$ \\
\midrule
\multirow{2}{*}{\shortstack[l]{\textbf{Text} \\ only}} & Ours & --- & \textbf{0.33} & 0.20 & 10.5 & \textbf{2.5} & --- & --- & --- & --- & --- \\
 & MusicGen & --- & 0.32 & \textbf{0.28} & \;\:\textbf{4.6} & 3.8 & --- & --- & --- & --- & --- \\ \hline
\multirow{2}{*}{\shortstack[l]{\textbf{Melody}\\(full)}} & Ours & \textbf{47.1}  & 0.33 & 0.22 & 10.8 & \textbf{2.5} & \textbf{82.6} & 0.33 & 0.19 & 11.2 & \textbf{2.0} \\
 & MusicGen & 41.3  & \textbf{0.34} & \textbf{0.29} & \;\:\textbf{5.7} & \textbf{2.5} & 55.2 & \textbf{0.34} & \textbf{0.28} & \;\:\textbf{6.2} & 2.8 \\\hline
 \multirow{2}{*}{\shortstack[l]{\textbf{Melody}\\(\sfrac{1}{2} prompt)}} & Ours & \textbf{46.7}  & 0.33 & 0.21 & 10.9 & 2.5 & \textbf{80.8} & 0.33 & 0.20 & 11.1 & \textbf{1.9} \\
 & MusicGen & 42.1  & \textbf{0.34} & \textbf{0.29} & \;\:\textbf{5.7} & \textbf{2.3} & 56.8 & \textbf{0.34} & \textbf{0.28} & \;\:\textbf{6.1 }& 2.3 \\
\bottomrule
\end{tabular}
\end{table*}

The 6-second  duration of our model can be restrictive for some real-world use cases.
Therefore, we capitalize on the inherent length-extrapolation ability of our fully convolutional model backbone, and experiment with 12 and 24 second-long generations (i.e., 2x and 4x the duration at training) using \expected{} melody controls.
The evaluation results are in Table~\ref{tab:extrap-results}.
We observe that both time-varying controllability and text control adherence are retained, but the overall audio realisticness, measured by FAD, somewhat degrades.
We verify this degradation via listening and note that the background noise level noticeably increases as we extrapolate duration.

\subsection{Benchmarking with MusicGen on Melody Control}

\begin{figure*}
    \centering
    \includegraphics[width=0.9\textwidth,
    , height=5.8cm]
    {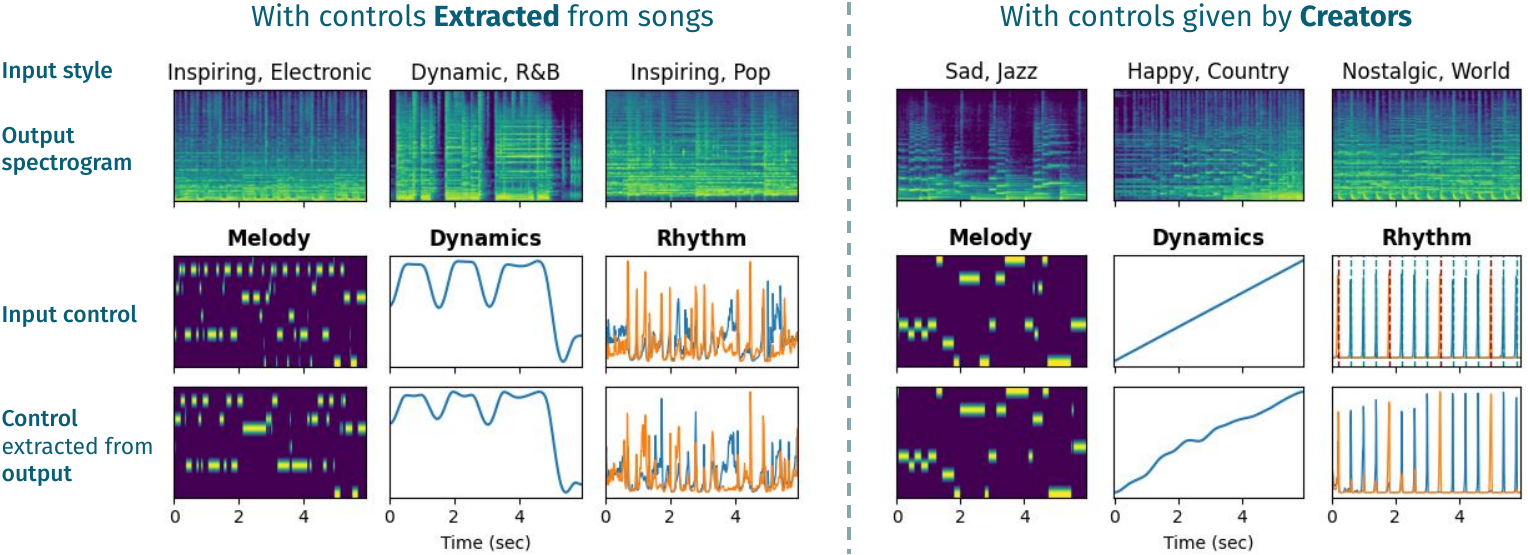}
     \caption{Examples of Music ControlNet generations given single time-varying controls.
     Our model faithfully follows all controls despite their non-obvious relationship with the spectrograms.
     Controls given by creators may be, e.g., simple melodies, drawn dynamics curves, and time-shifted/stretched rhythm templates as shown here.
     (Colors in rhythm control represent {\color{NavyBlue}{beat}} / {\color{Orange}{downbeat}} probabilities.
     Dashed lines in the creator rhythm control are beat/downbeat timestamps that can be used to sync beats as desired.)}\label{fig:single-ctrls}
     \vspace{-2mm}
\end{figure*}

We compare our model trained with melody, dynamics, and rhythm controls to the 1.5B-parameter MusicGen~\pcite{copet2023simple} model trained only with melody control. 
We use the MusicGen model in three scenarios: 
(i)~\textit{text-only generation}, where we do not pass in melodies, 
(ii)~\textit{full melody control}, where we pass in melodies that are as long as generation length, and 
(iii)~\textit{1/2 prompt melody control}, where the melodies passed in are half length.
For our model, these scenarios are achieved by omitted, partially-specified, or full melody control. 

As MusicGen support free-form text control while our model does not,
in this experiment, we use both the \textbf{MusicCaps} and \textbf{MusicCaps+ChatGPT} datasets.
Both datasets contains the same audio, but the \textbf{MusicCaps+ChatGPT} dataset has the text descriptions converted into genre \& mood tags by ChatGPT.
The ChatGPT-converted tags are then used in two ways: as the global style input to our model, and as text input when computing CLAP scores.
That is, we have two versions of CLAP when comparing our model to MusicGen, namely, \textbf{CLAP}$_{\text{text}}$, which measures CLAP with (original free text, generation audio) tuples, and \textbf{CLAP}$_{\text{tag}}$ (i.e., the CLAP metric used in previous experiments), which only allows converted tags as text input to both MusicGen (written as text, e.g., ``\textit{An audio of happy jazz music}'') and our model, and measures CLAP with (converted tags, generation audio) tuples.
We also compute two versions of FAD scores, one using MusicCaps as the reference set (i.e., \textbf{FAD}$_{\text{MCaps}}$) and the other using our in-domain test set as the reference (i.e., \textbf{FAD}$_{\text{Ours}}$).
We generate 10-second long outputs to be consistent with the MusicCaps dataset and evaluation protocol.

We consider both \textit{\extracted{}} and \textit{\expected{}} melody controls in this comparison.
As shown in Table~\ref{tab:vs-musicgen}, we find our proposed work responds more precisely to the melody control, particularly on \expected{} melodies, where our model is as much as 49\% relatively more faithful to the control.
In terms of text control adherence, when the text input is restricted to the converted mood \& genre tags (i.e.,  the \textbf{CLAP}$_{\text{tag}}$ metric), our model is comparable to MusicGen.
On overall audio realisticness, as our model is much smaller than MusicGen, and trained on a much more restricted domain of data,
it is unsurprising that it scores a worse FAD when using MusicCaps recordings.
Moreover, we note that many examples in the MusicCaps datast are, in fact, low-quality audio recordings and/or contain vocals which our model never sees during training, which 
may render $\textbf{FAD}_{\text{MCaps}}$ biased against our model.
Finally, we note when the reference set is our in-domain test set audios (i.e., \textbf{FAD}$_{\text{ours}}$), we are competitive to or somewhat better than MusicGen.


\begin{figure*}
    \centering
    \includegraphics[width=0.85\textwidth]{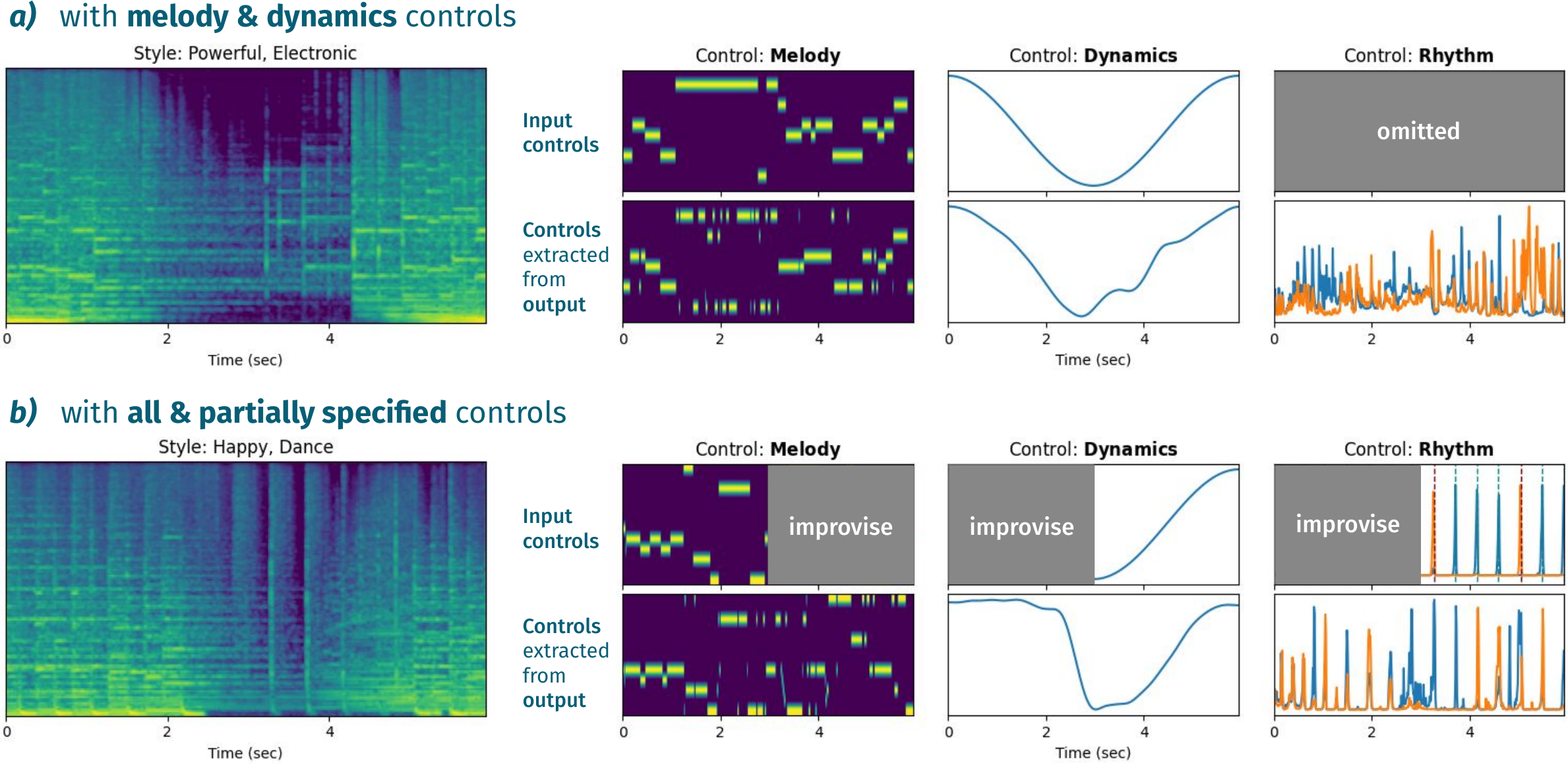}
    \caption{Music ControlNet generations with multiple and/or partially-specified controls given by \textit{creators}. All controls are honored when enforced, demonstrating the composability of our controls. In uncontrolled segments, the generations exhibits consistent style and musical creativity.}
    \label{fig:multi-and-dontcare}
    \vspace{-2mm}
\end{figure*}

\subsection{Qualitative Analysis of Generations}
In Fig.~\ref{fig:single-ctrls}, we show generation outputs with each of the proposed controls, i.e., melody, dynamics, or rhythm, either \textit{\extracted{}} or \textit{\expected{}}.
Concentrating first on the \extracted{} controls (Fig.~\ref{fig:single-ctrls}, left half), all of the three control signals are closely followed by our model even with their different dimensions and relationships w.r.t.~the spectrogram.
Moving on to the \expected{} controls (Fig.~\ref{fig:single-ctrls}, right half), the controls are almost perfectly reflected despite some of them (i.e., melody \& dynamics) being out-of-domain from training data.
Moreover, our approach is able to wield musical creativity even though the \expected{} controls are much simpler than \extracted{} ones.
For example, visible from the output spectrograms given melody or dynamics controls, 
our model generates music with varying texture and rhythmic patterns, rather than simply replicating the monophonic melody, or changing the volume of a single note to match the increasing dynamics.

Fig.~\ref{fig:multi-and-dontcare} displays generations using multiple \expected{} controls, specifically, with \textit{a)}~full melody \& dynamics controls simultaneously enforced, and \textit{b)}~all three controls with partially-specified spans, which simulate the creator intent: ``\textit{I want the music to start with my signature melody, and have it intensifying at the end with beats synchronized to my video scene cuts to engage my audience.}''
Example \textit{a)} verifies the composability of \expected{} controls (as opposed to \extracted{} ones, which has been examined in Table~\ref{tab:single-vs-multi}) as both controls are respected by the model.
Example \textit{b)} demonstrates effective control even when controls signals are partially specified, and the capability to generate cohesive music (i.e., the output spectrogram contains no visible borders) when both controlled and uncontrolled spans are present.

%% file: sections/related.tex
\section{Related Work}\label{sec:related}

\subsection{Text-to-music Generation} 
Music ControlNet builds on a recent body of work on text-to-music, where the goal is to generate music audio conditioned on text descriptions or categories~\citep{dhariwal2020jukebox,forsgren2022riffusion,agostinelli2023musiclm,huang2023noise2music,liu2023audioldm,chen2023musicldm}. 
This line of research is bifurcated into two broad methodological branches which build on advances in natural language processing and computer vision respectively:  
(i)~using LLMs to model tokens from learned audio codecs as proposed in~\citep{van2017vqvae,dieleman2018challenge}, and 
(ii)~using (latent) diffusion to model image-like spectrograms. 
We explore diffusion to leverage strong inductive biases developed for spatial control.

\subsection{Time-varying Controls for Music Generation}
Our approach is related to generating music audio from time-varying control.
A contemporaneous work is~\citep{lin2023content}, which focuses on a similar goal to ours, but is built on pretrained large language models~(LLMs) instead of diffusion models.
Work on style transfer includes methods to convert musical recordings in one style to another while preserving underlying symbolic music~\citep{mor2018universal,huang2018timbretron,engel2020ddsp,caillon2021rave}. 
Other work explores directly synthesizing symbolic music (e.g., MIDI) into audio~\citep{hawthorne2018enabling,hawthorne2022multi}. 
Both approaches require training individual models per style rather than leveraging text control for style, and needs complete musical inputs rather than simpler controls we explore here. 
More recently, \cite{wu2022jukedrummer,donahue2023singsong,garcia2023vampnet} generate music in broad styles with time-varying control but target tasks with stronger conditions like musical accompaniment or variation generation, which are different applications than ours.
Another body of research~\citep{chen2020music, tan2020music, dai2021controllable, wu2023musemorphose} explores time-varying controls for symbolic-domain music generation, i.e., modeling sheet music or MIDI events.
The controls considered in these works are of coarser time scale, e.g., at the measure or phrase level, while our approach offers precise control down to the frame level.

\subsection{Unconditional Music Generation}
Our work on controllable music audio generation builds on earlier work on unconditional generative modeling of audio.  
Early approaches explored directly modeling audio waveforms~\citep{oord2016wavenet,kalchbrenner2018efficient,donahue2018adversarial}. 
More recent work~\citep{van2017vqvae,dieleman2018challenge,hawthorne2022general,borsos2023audiolm} favors hierarchical approaches like those we consider there.

%% file: sections/conclusion.tex
\section{Conclusions}
We proposed Music ControlNet, a framework that enables creators to harness music generation with precise, multiple time-varying controls.
We demonstrate our framework via melody, dynamics, and rhythm control signals, which are all basic elements in music and complement with each other well.
We find that our framework and control signals not only enables any combination of controls, fully- or partially-specified in time, but also generalizes well to controls we envision creators would employ.

Our work paves a number of promising avenues for future research.
First, beyond melody, dynamics, and rhythm controls, several additional musical features could be employed such as chord estimation for harmony control, multi-track pitch transcription, instrument classification, or even more abstract controls like emotion and tension.
Second, as the set of musical controls becomes large, generating control presets based on text, speech, or video inputs could make controllable music generation systems more approachable to a wide range of content creators.
Last but not least, addressing the domain gap between \extracted{} and \expected{} controls via, e.g., adversarial approaches~\citep{kim2022refining}, could further enhance the musical quality of generations under \expected{} controls.

%% file: sections/ethics.tex
\section{Ethics Statement}
Music generation is poised to upend longstanding norms around how music is created and by whom. 
On the one hand, this presents an opportunity to increase the accessibility of musical expression, but on the other hand, existing musicians may be forced to compete against generated music. 
While we acknowledge our work carries some risk, we sharply focus on improving control methods so as to directly offer musicians more creative agency during the generation process. 
Other potential risks surround the inclusion of singing voice, accidental imitation of artists without their consent, and other unforeseen ethical issues, so we use licensed instrumental music for training and melodies extracted from our training data or public domain melodies we recorded ourselves for inference. For evaluation, we do use the MusicCaps dataset~\pcite{agostinelli2023musiclm} as it is standard in recent text-to-music generation literature.